\documentclass[12pt,a4paper,oneside]{article}
\usepackage[T1]{fontenc}

\usepackage[utf8]{inputenc}
\usepackage{lmodern}

\usepackage{geometry} 
\usepackage{color}
\usepackage{amsmath} 
\font\kapitaliki=plcsc10 

\title{The Shortest Confidence Interval\\
for the Ratio of Quantiles of the Dagum Distribution\\}
\date{}

\author{
Alina J\c{e}drzejczak\\
Dorota Pekasiewicz\\
University of {\L}{\'o}d{\'z} (Poland)\\
e-mail: alina.jedrzejczak@uni.lodz.pl\\		
e-mail: dorota.pekasiewicz@uni.lodz.pl\\
\\
Wojciech Zieli\'{n}ski\\
Warsaw University of Life Sciences (Poland)\\
e-mail: wojciech$\_$zielinski@sggw.pl\\	}	

\begin{document}
\maketitle
\begin{abstract}
	{\noindent Jędrzejczak et al. (2018) constructed a confidence interval for a ratio of quantiles coming from the Dagum distribution, which is frequently applied as a theoretical model in numerous income distribution analyses.  The proposed interval is symmetric with respect to the ratio of sample quantiles, which result may be unsatisfactory in many practical applications. The search for a confidence interval with a smaller length led to the derivation of the shortest interval with the ends being asymmetric relative to the ratio of sample quantiles. In the paper, the existence of the shortest confidence interval is shown and the  method of obtaining such an interval is presented. The results of the calculation show a reduction in the length of the confidence intervals by several percent in relation to the symmetric confidence interval. \\
	}
\end{abstract}
     {\kapitaliki Keywords}:  quantile share ratio, ratio of quintiles, the shortest confidence interval, Dagum distribution
	
	\section{Introduction}
	
	One of the measures of income distribution inequality is  income quintile share ratio, also called  S80/S20 ratio (Eurostat Regional Yearbook, 2016). It is calculated as the ratio of the fourth quantile of an income distribution to the first one, i.e. income quintile share ratio can be defined as
	$$r_{0.2,0.8}=\frac{F^{-1}(0.8)}{F^{-1}(0.2)},$$
	where $F$ denotes the cumulative distribution function of income. \\
	\indent Natural point estimators of the ratio $r_{0.2,0.8}$ can easily be obtained as ratios of the corresponding sample quintiles. Statistical properties of such estimators are dependent on the form of quantile estimators which are implemented  (see: Jędrzejczak and Pekasiewicz, 2018).
	In addition to point estimation,  in the analysis of income distribution,  the interval estimation of different inequality measures is also considered. In Greselin and Pasquazzi (2009, 2010),
	parametric and non-parametric Dagum confidence intervals for Gini and new Zenga inequality measures were derived and compared. Jędrzejczak et al. (2018) constructed a confidence interval for the ratios of quantiles, assuming that the Dagum distribution as an income distribution model. Also in this paper we confine ourselves to the Dagum (1977) distribution
	as a probabilistic model for a population income. This distribution has widely been applied in income inequality analysis in many countries all over the world as it is very flexible and usually well fitted to empirical distributions in different divisions.
		Here, we consider a more general set-up, namely a confidence interval for a ratio of  $\alpha$ and $\beta$ quantiles is taken into regard.  Jedrzejczak et al. (2018) constructed a symmetric confidence interval, i.e. the confidence interval  for which the  risks of underestimation and overestimation are the same. The current study is dedicated to the problem of construction of the shortest confidence interval.  \\
	\indent 	In the second section the Dagum distribution is presented while in the third section the shortest confidence interval is derived. Unfortunately closed formulae turned out to be not available. Some numerical results are given in the fourth section. In the last section conclusions and final remarks are presented, as well as the suggestions of future research topics
		
	\section {The Dagum Distribution}
	The Dagum distribution is often used in the analysis of personal (or household) income and wages, as it is usually well fitted to empirical distributions in different countries. It can also be successfully applied  for different  subpopulations obtained by means of splitting up the overall sample by socio-economic group, region, gender or family type  (Jędrzejczak and  Pekasiewicz 2018, Pekasiewicz and Jędrzejczak 2017). The estimates of its parameters are utilized to assess many important income distribution characteristics, including numerous variability, inequality, poverty and wealth measures, as well as concentration curves.\\
	\indent Consider a Dagum distribution with parameters $a>0$, $v>0$ and $\lambda>0$. Its cumulative distribution distribution (CDF) and probability density function (PDF) are as follows
	$$F_{a, v, \lambda}(x)=\left(1+\left(\frac{x}{\lambda}\right)^{-v}\right)^{-a}\hbox{ for }x>0$$
	and
	$$f_{a, v, \lambda}(x)=\frac{av}{\lambda}\left(\frac{x}{\lambda}\right)^{av-1}\left(1+\left(\frac{x}{\lambda}\right)^v\right)^{-a-1}\hbox{ for }x>0.$$
	Its quantile function equals
	$$Q_{a, v, \lambda}(q)=\lambda\left(q^{-1/a}-1 \right)^{-1/v}\hbox{ for }0<q<1.$$
	
	\section{The Shortest Confidence Interval}
	Among income distribution  characteristics the important role play the measures based on quantiles. Simple dispersion ratios, defined as the ratios of the income of the richest quantile over that of the poorest quantile, usually utilize deciles and quintiles, but in principle, any quantile of income distribution can be used. A version of the decile dispersion ratio using the ratio of the 90ty over the 40th percentile (the ratio of the richest 10\%  divided by the poorest 40\%’s)  which has recently become popular is the so called Palma Ratio (Palma, 2011) Another popular inequality measure based on quantiles (deciles) is the coefficient of maximum  equalisation, also known as the Schutz index or the Pietra ratio.\\
	\indent Let $0<\alpha<\beta<1$ be given numbers. We are interested in estimation of ratio of quantiles
	$$r_{\alpha,\beta}=\frac{Q_{a, v, \lambda}(\beta)}{Q_{a, v, \lambda}(\alpha)}=\frac{\left(q^{-1/\beta}-1 \right)^{-1/v}}{\left(q^{-1/\alpha}-1 \right)^{-1/v}}.$$
	Since we are interested in the estimation of the ratio $r_{\alpha,\beta}$ of quantiles, we reparametrize the considered model. It can be seen that
	$$v=\frac{\log\left(\frac{\alpha^{-1/a}-1}{\beta^{-1/a}-1}\right)}{\log r_{\alpha,\beta}}.$$
	\indent The CDF of the Dagum distribution may be written in the following form
	$$F_{a, r_{\alpha,\beta}, \lambda}(x)=\left(1+\left(\frac{x}{\lambda}\right)^{-\frac{\log\left(\frac{\alpha^{-1/a}-1}{\beta^{-1/a}-1}\right)}{\log r_{\alpha,\beta}}}\right)^{-a}$$
	for $x>0$ and  $a>0$, $r_{\alpha,\beta}>0$ and $\lambda>0$.
	
	Let $X_1,\ldots,X_n$ be a sample and let $X_{1:n}\leq X_{2:n}\leq\cdots\leq X_{n:n}$ be order statistics. Let $r^*_{\alpha,\beta}=\frac{X_{\lfloor\beta n\rfloor+1:n}}{X_{\lfloor\alpha n\rfloor+1:n}}$ (here $\lfloor x\rfloor$ denotes the greatest integer not greater than $x$) be an observed quantile ratio. It is assumed that the sample size $n$ is large, i.e. it is assumed that $n\to\infty$.
	
	From David and Nagaraja (2003) and Serfling (1999) it follows that $r^*_{\alpha,\beta}$ is strongly consistent estimator of $r_{\alpha,\beta}$, for all $a,v,\lambda$. Also, it follows (Serfling 1999, th. 2.3.3; David and Nagaraja 2003, th. 10.3 and application of Delta method, see e.g. Greene 2003, p. 913) that for $0<\alpha<\beta<1$ the estimator $r^*_{\alpha,\beta}$ is asymptotically normally distributed, i.e.
	$$\sqrt{n}\left(r^*_{\alpha,\beta}-r_{\alpha,\beta}\right)\to r_{\alpha,\beta}N\left(0,w^2(a)\right),$$
	where
	$$\begin{aligned}
	w^2(a)=&\frac{1}{(av)^2}
	\left(\frac{1-\beta}{\beta}\frac{1}{(1-\beta^\frac{1}{a})^2}+\frac{1-\alpha}{\alpha}\frac{1}{(1-\alpha^\frac{1}{a})^2}-2\frac{1-\beta}{\beta}\frac{1}{(1-\alpha^\frac{1}{a})(1-\beta^\frac{1}{a})}\right)\\
	=&\left(\frac{1}{a\log\left(\frac{\alpha^{-1/a}-1}{\beta^{-1/a}-1}\right)}\right)^2\\
	&\hskip2em\cdot\left(\frac{1-\beta}{\beta}\frac{1}{(1-\beta^\frac{1}{a})^2}+\frac{1-\alpha}{\alpha}\frac{1}{(1-\alpha^\frac{1}{a})^2}-2\frac{1-\beta}{\beta}\frac{1}{(1-\alpha^\frac{1}{a})(1-\beta^\frac{1}{a})}\right)
	\end{aligned}$$
	(for theoretical details see Jędrzejczak et al. (2018)).
	
	Let $\delta$ be the given confidence level. We have (the scale parameter $\lambda$ is omitted)
	$$P_{r,a}\left\{u_{\delta_1-\delta}\leq\sqrt{n}\frac{r^*_{\alpha,\beta}-r_{\alpha,\beta}}{r_{\alpha,\beta}\log r_{\alpha,\beta}w(a)}\leq u_{\delta_1}\right\}=\delta,$$
	where $\delta\leq\delta_1\leq1$ and $u_{\gamma}$ is the $\gamma$-quantile of $N(0,1)$ distribution.
	
	Let
	$$EoCI\left(\gamma\right)=\frac{r^*_{\alpha,\beta}z_{\gamma}(a)}{W\left(r^*_{\alpha,\beta}z_{\gamma}(a)\exp\left(z_{\gamma}(a)\right)\right)},$$
	where  $z_{\gamma}(a)=\frac{\sqrt{n}}{u_{\gamma}w(a)}$ and $W(\cdot)$ is the Lambert $W$ function (see Appendix~2). The confidence interval for $r_{\alpha,\beta}$ at the confidence level $\delta$ has the form
	$$\left(EoCI\left(\delta_1\right);\ EoCI\left(\delta_1-\delta\right)\right).$$

	The confidence interval with $\delta_1=(1+\delta)/2$ is the standard one, i.e.
	$$\left(EoCI\left(\frac{1+\delta}{2}\right);\ EoCI\left(\frac{1-\delta}{2}\right)\right).$$
	The length of the confidence interval is a function of $\delta_1$:
	$$L\left(\delta_1\right)=EoCI\left(\delta_1-\delta\right)-EoCI\left(\delta_1\right).$$
	We want to minimize $L\left(\delta_1\right)$ with respect to $\delta_1$.
	
	\bigskip
	
	{\bf Lemma.} $\frac{W(z)}{W(az)}$ is decreasing for $a>1$; is constant for $a=1$; is increasing for $0<a<1$.
	
	\medskip
	{\it Proof.} To obtain the thesis it is enough to observe that function $W(z)$ is increasing for $z>e^{-1}$ and it is convex. These properties as well as other interesting properties of the Lambert $W$ function may be found in Corless et al. (1996).
	
	\bigskip
	
	{\bf Theorem.} There exists $\delta_1$ which minimizes $L\left(\delta_1\right)$.
	
	\medskip
	{\it Proof.} If $\gamma\in(0,1)$ increases then $z_{\gamma}\exp(z_{\gamma})$ decreases. Since $r^*_{\alpha,\beta}>1$ and $z=W(ze^z)$ hence $EoCI(\gamma)$ decreases. We have:
	
	if $\delta_1\searrow\delta$ then $EoCI\left(\delta_1\right)\to EoCI\left(\delta\right)<\infty$ and $EoCI\left(\delta_1-\delta\right)\nearrow+\infty$;
	
	if $\delta_1\nearrow1$ then $EoCI\left(\delta_1\right)\searrow -\infty$ and $EoCI\left(\delta_1-\delta\right)\to EoCI\left(1-\delta\right)<\infty$.
	
	Hence
	$$\delta_1\searrow\delta\Rightarrow L\left(\delta_1\right)\nearrow+\infty\ \hbox{and}\ \delta_1\nearrow1\Rightarrow L\left(\delta_1\right)\nearrow+\infty.$$
	From continuity of $EoCI(\cdot)$ we obtain the thesis.
	
	\medskip

	Note that for $\delta_1=\delta$ and $\delta_1=1$ we obtain one sided confidence intervals.
	
	\bigskip
	
	\section{Numerical results}

	The analytical form of $\delta_1$ minimizing the length of the confidence interval for quantile ratios of the Dagum distribution cannot be obtained but it can be found numerically.  In Appendix 1 there is given a short code in R-project language for finding the minimal length confidence interval. \\
	\indent Exemplary numerical results are given in Tables 1 and 2 for $n=1000$ and $\delta=0.95$. In particular, Table 1 summarizes the results of the calculations obtained  for $\alpha=0.2=1-\beta$ while in Table~2 the results for $\alpha=0.1=1-\beta$ are presented. In columns entitled  {\sl short} and {\sl standard} there are given appropriate interval lengths and the last column contain the corresponding length reductions which can be considered the precision gains obtained by means of the proposed estimation method. Note that $1-\delta_1$ is the risk of underestimation while $\delta_1-\delta$ is the risk of overestimation (for standard confidence interval both probabilities are equal to $(1-\delta)/2=0.025$).
	
	
	$$\vbox{\tabskip=2em minus1.9em\offinterlineskip\halign to\hsize{
			\strut\hfil$#$\hfil&#\vrule&\hfil$#$\hfil&#\vrule\hskip2pt\vrule&&\hfil$#$\hfil&#\vrule\cr
			\multispan{13} {\bf Table 1.} Results for $\alpha=0.2=1-\beta$.\hfill\cr\noalign{\vskip2pt}
			a&&r&&\delta_1-\delta&&1-\delta_1&&\hbox{short}&&\hbox{standard}&&\hbox{reduction}\cr\noalign{\hrule}
			0.1&&2.0&&0.03339&&0.01661&&0.241322&&0.244047&&1.117\%\cr
			0.1&&3.0&&0.03527&&0.01473&&0.577946&&0.588114&&1.729\%\cr
			0.1&&4.0&&0.03646&&0.01354&&0.978297&&1.000800&&2.248\%\cr
            0.1&&5.0&&0.03740&&0.01260&&1.427151&&1.466759&&2.700\%\cr
            0.1&&6.0&&0.03813&&0.01187&&1.915428&&1.976761&&3.103\%\cr
			\noalign{\hrule}
			0.5&&2.0&&0.03217&&0.01783&&0.202939&&0.204560&&0.792\%\cr
			0.5&&3.0&&0.03374&&0.01626&&0.484962&&0.490972&&1.224\%\cr
			0.5&&4.0&&0.03486&&0.01514&&0.819397&&0.832625&&1.589\%\cr
            0.5&&5.0&&0.03569&&0.01431&&1.193435&&1.216616&&1.905\%\cr
            0.5&&6.0&&0.03635&&0.01365&&1.599462&&1.635214&&2.186\%\cr
			\noalign{\hrule}
			1.0&&2.0&&0.03176&&0.01824&&0.191330&&0.192689&&0.705\%\cr
			1.0&&3.0&&0.03329&&0.01671&&0.456954&&0.461982&&1.088\%\cr
			1.0&&4.0&&0.03436&&0.01564&&0.771701&&0.782753&&1.412\%\cr
            1.0&&5.0&&0.03515&&0.01485&&1.123495&&1.142840&&1.693\%\cr
            1.0&&6.0&&0.03578&&0.01422&&1.505166&&1.534971&&1.942\%\cr
 	}}$$

	$$\vbox{\tabskip=2em minus1.9em\offinterlineskip\halign to\hsize{
			\strut\hfil$#$\hfil&#\vrule&\hfil$#$\hfil&#\vrule\hskip2pt\vrule&&\hfil$#$\hfil&#\vrule\cr
			\multispan{13} {\bf Table 2.} Results for $\alpha=0.1=1-\beta$.\hfill\cr\noalign{\vskip2pt}
			a&&r&&\delta_1-\delta&&1-\delta_1&&\hbox{short}&&\hbox{standard}&&\hbox{reduction}\cr\noalign{\hrule}
			0.1&&2.0&&0.03308&&0.01692&&0.231886&&0.234303&&1.032\%\cr
			0.1&&3.0&&0.03490&&0.01510&&0.555027&&0.564033&&1.597\%\cr
			0.1&&4.0&&0.03608&&0.01392&&0.939046&&0.958947&&2.075\%\cr
            0.1&&5.0&&0.03699&&0.01301&&1.369310&&1.404301&&2.492\%\cr
            0.1&&6.0&&0.03770&&0.01230&&1.837099&&1.891224&&2.862\%\cr
			\noalign{\hrule}
			0.5&&2.0&&0.03176&&0.01824&&0.190784&&0.192131&&0.701\%\cr
			0.5&&3.0&&0.03329&&0.01671&&0.455637&&0.460622&&1.082\%\cr
			0.5&&4.0&&0.03436&&0.01564&&0.769406&&0.780416&&1.404\%\cr
            0.5&&5.0&&0.03512&&0.01488&&1.120211&&1.139388&&1.683\%\cr
            0.5&&6.0&&0.03575&&0.01425&&1.500741&&1.530285&&1.931\%\cr
			\noalign{\hrule}
			1.0&&2.0&&0.03120&&0.01880&&0.174992&&0.176031&&0.591\%\cr
			1.0&&3.0&&0.03264&&0.01737&&0.417617&&0.421456&&0.911\%\cr
			1.0&&4.0&&0.03364&&0.01636&&0.704826&&0.713249&&1.181\%\cr
            1.0&&5.0&&0.03436&&0.01564&&1.025577&&1.040298&&1.415\%\cr
            1.0&&6.0&&0.03496&&0.01504&&1.373322&&1.395974&&1.623\%\cr
	}}$$
	
	\section{Conclusions}
	
	One of the crucial problems in socio-economic research is estimation of  income distribution inequality which can be evaluated, among others, by the ratio of appropriate quantiles of an income distribution. Such an approach is very convenient for practitioners, as the inequality measures based on quantiles are easy to obtain and have straightforward economic interpretation. In Jędrzejczak et al. (2018) an asymptotic confidence interval for a ratio of quantiles was constructed. \\
	\indent In this paper we constructed the shortest confidence interval.  We confined ourselves to the Dagum distribution which was assumed as an underlying income distribution model throughout the paper.   It was just because this distribution presents good statistical properties required for income distribution models and is widely applied  in numerous empirical analyses. The confidence interval we constructed is asymptotic but in the real-world  experiments on income and wage distributions thousands of data are available. Numerous simulation studies performed in Zielinski et al. (2018) revealed that under the Dagum model the sample size n = 1000 is large enough to do asymptotics. \\
	\indent The empirical analysis of the lengths of c.i.  for quintile and decile ratios confirmed a reduction in the length of the proposed confidence interval by several percent with respect to the symmetric one. It is worth noting that the observed  length reduction has strictly been related to the statistical characteristics of the Dagum distribution, namely its dispersion and inequality. The greater income inequality is observed the smaller the precision of interval estimation can be expected and the more reduction you can get due to the new approach.  Therefore,  the proposed shortest confidence interval can be applied in various income, wage and expenditure analysis, wherever we can successfully utilize the Dagum distribution. Because nowadays it is easy to calculate the shortest c.i. hence these intervals can be recommended for practical use.
	In the future, further investigations on confidence intervals for income inequality and poverty measures, involving different probability distributions, seem useful.

	\section*{References} 
	
	\font\kapitaliki=plcsc10
	\def\art#1#2#3#4#5{\noindent\hangindent=0.5truecm \hangafter=1 {\kapitaliki #1}\ (#2):\ ``#3,''\ {\it #4}, #5.\par}
	\def\book#1#2#3#4{\noindent\hangindent=0.5truecm \hangafter=1 {\kapitaliki #1}\ (#2):\ ``#3,''\ {#4}.\par}

	\art{Corless, R. M., Gonnet, G. H., Hare, D. E. G., Jeffrey, D. J. and Knuth D. E.}{1996}{On the LambertW function}{Advanced in Computational Mathematics}{5(1), 329–359}
	
	\art{Dagum, C.}{1977}{A New Model of Personal Income Distribution: Specification and Estimation}{Economie Appliquee}{30, 413-437}
	
	\book{David, H. A. and Nagaraja, H. N.}{2003}{Order Statistics, Third Edition}{John Wiley \& Sons, Inc}
	
	
	
	
	
	\noindent\hangindent=0.5truecm \hangafter=1 {\kapitaliki Eurostat}\ (2016):\ ``The Eurostat regional yearbook ISBN: 978-92-79-60090-6, ISSN: 2363\=1716, doi: 10.2785/29084, cat. number: KS-HA-16-001-EN-N. (http://ec.europa.eu/eurostat/statistics-explained/index.php/Glossary:Income$\_$quintile$\_$share$\_$ratio).''\par
	
	\book{Greene, W. H.}{2003}{Econometric Analysis (5th ed.)}{Prentice Hall}
	
	\art{Greselin, F. and Pasquazzi, L.}{2009}{Asymptotic Confidence Intervals for a New Inequality Measure}{Communications in Statistics - Simulation and Computation,TaylorFrancis}{38(8), 1742-1756}
	
	\art{Greselin, F. and Pasquazzi, L.}{2010}{Dagum Confidence Intervals for Inequality Measures}{Proceedings of the 45th Scientific Meeting of the Italian Statistical Society, Cleup Eds., Padua}{https://www.academia.edu/17707359/Dagum confidence intervals for inequality measures}	
	
	
	\art{Jędrzejczak, A. and Pekasiewicz D.}{2018}{Properties of Selected Inequality Measures Based on Quantiles and their Application to the Analysis of Income Distribution in Poland by Macroregion}{Argumenta Oeconomica Cracoviensia}{18, 51-67}

	\art{Jędrzejczak, A., Pekasiewicz, D. and Zieliński, W.}{2018}{Confidence Interval for Quantile Ratio of the Dagum Distribution}{REVSTAT-Statistical Journal}{https://ine.pt/revstat/pdf/ ConfidenceIntervalQuantileRatioDagum.pdf}
			
    \art{Palma, J. G.}{2011}{Homogeneous Middles vs. Heterogeneous Tails, and the End of the 'Inverted-U': the Share of the Rich is what it's all about}{Cambridge Working Papers in Economics (CWPE) 1111}{}

    \art{Pekasiewicz, D. and Jędrzejczak, A.}{2017}{Application of the Measures Based on Quantiles to the Analysis of Income Inequality and Poverty in Poland by Socio-Economic Group}{35th International Conference Mathematical Methods in 	Economics}{532-537}

  	\book{Serfling, R. J.}{1980}{Approximation Theorems of Mathematical Statistics}{John Wiley \& Sons}
	
	\art{Zieliński, W., Jędrzejczak, A. and Pekasiewicz, D.}{2018}{Estimation of quantile ratios of the Dagum distribution}{The $12^{th}$
	Professor Aleksander Zeliaś International Conference on Modelling and Forecasting of Socio-Economic Phenomena}{12, 603-611, \\ DOI: 10.14659/SEMF.2018.01.61}
	
	\vfill\eject
	
	\section*{Appendix 1}
	
	An exemplary R code for calculating the confidence interval is enclosed.
	\bigskip
	
	\begingroup
	\parindent0pt
	\baselineskip=9pt
	\font\ttt=pltt8
	\ttt
	\def\hs{\hskip0.5em}
	\obeylines
	
	alpha=0.2 \#input alpha
	beta=0.8 \#input beta
	n=1000 \#input n
	rsample=2.5 \#input estimated r
	asample=0.1 \#input estimated shape parameter a
	lev=0.95 \#input confidence level
	stlev=(1-lev)/2
	kryt=function(s)$\{$qnorm(s,0,1)$\}$
	
	sigma2=function(A,B,aa)$\{$(aa*log((A\^{}(-1/aa)-1)/(B\^{}(-1/aa)-1)))\^{}(-2)*((1-B)/B/(1-B\^{}(1/aa))\^{}2
	\hs +(1-A)/A/(1-A\^{}(1/aa))\^{}2-2*(1-B)/B/((1-B\^{}(1/aa))*(1-A\^{}(1/aa))))$\}$
	
	TheEnd=function(pr,nn,A,B,aa,kryt)$\{$uniroot(function(pop) sqrt(nn)*(pr-pop)
	\hs -kryt*sqrt(sigma2(A,B,aa))*pop*log(pop), lower = 1, upper = 10, tol = 1e-20)\$root$\}$
	
	Leng=function(pr,nn,A,B,aa,s)
	\hs $\{$TheEnd(pr,nn,A,B,aa,kryt(s))-TheEnd(pr,nn,A,B,aa,kryt(lev+s))$\}$
	\hs \#s -risk of overestimation to be optimized
	
	FindMinimumLeng=function(pr,nn,A,B,aa,ll)$\{$optimize(Leng,interval=c(0,1-ll), pr=pr, nn=nn,
	\hs A=A, B=B, aa=aa, tol=1e-20)\$minimum$\}$
	
	ss=FindMinimumLeng(rsample,n,alpha,beta,asample,lev)
	
	print(c("\ risk of overestimation:",ss),quote = FALSE)
	print(c("\ length of the shortest confidence interval:",
	\hs Leng(rsample,n,alpha,beta,asample,ss)),quote = FALSE)
	print(c("\ length of the standard confidence interval:",
	\hs Leng(rsample,n,alpha,beta,asample,stlev)),quote = FALSE)
	print(c("\ length reduction:",100*
	\hs (1-Leng(rsample,n,alpha,beta,asample,ss)/Leng(rsample,n,alpha,beta,asample,stlev)),"\%"),
	\hs quote = FALSE)
	print(c("\ standard c.i.:",TheEnd(rsample,n,alpha,beta,asample,kryt(lev+stlev)),
	\hs TheEnd(rsample,n,alpha,beta,asample,kryt(stlev))), quote = FALSE)
	print(c("\ shortest c.i.:",TheEnd(rsample,n,alpha,beta,asample,kryt(lev+ss)),
	\hs TheEnd(rsample,n,alpha,beta,asample,kryt(ss))), quote = FALSE)
	
	\endgroup
	
	\vfill\eject
	
	\section*{Appendix 2}
	
	Lambert function $W(\cdot)$ is defined as a solution with the respect to $t$ of the equation $$te^t=z\Rightarrow t=W(z).$$
	It is seen that
	$$W(z)e^{W(z)}=z \Rightarrow  W(z)=\ln\left(\frac{z}{W(z)}\right)\Rightarrow z=\frac{z}{W(z)}\ln\left(\frac{z}{W(z)}\right).$$
	Since the solution with respect to $r$ of the equation $r\ln r=z$ is $r=\frac{z}{W(z)}$, hence
	$$
	A\frac{x-r}{r\ln r}=1\Rightarrow Ax=r(\ln r+A)\Rightarrow e^AAx=\left(re^A\right)\ln\left(re^A\right)\Rightarrow r=\frac{Ax}{W(Axe^A)}.
	$$
	Application of the above to the equation
	$$\sqrt{n}\frac{r^*_{\alpha,\beta}-r_{\alpha,\beta}}{w(a)r_{\alpha,\beta}\log r_{\alpha,\beta}}= u$$
	gives the confidence interval for the ratio $r_{\alpha,\beta}$.
	
\end{document}